\documentstyle[12pt]{article}
\title{Bond Orientational Order, Molecular Motion \\ and Free Energy \\
of High Density DNA Mesophases}
\author{{R. Podgornik \thanks{\mbox{On leave from J.Stefan Institute,
Ljubljana, Slovenia.}}, H.H. Strey, K. Gawrisch$^{\ddagger}$, }\\
{D.C. Rau$^{\sharp}$, A. Rupprecht$^{\star}$ and V.A.
Parsegian \thanks{To whom correspondence should be addressed.}}
\\ ~
\\ {Laboratory of Structural Biology DCRT}
\\ $^{\sharp}${Division of Intramural Research NIDDK}
\\ $^{\ddagger}${Division of Intramural Clinical and Biological
Research NIAAA}
\\ {National Institutes of Health, Bethesda, MD 20892}
\\ ~
\\ $^{\star}$ Physical Chemistry Department, Stockholm University, Sweden}

\begin{document}
\maketitle
\begin{titlepage}

\begin{abstract}
\baselineskip=15pt

By equilibrating condensed DNA arrays against reservoirs of known
osmotic stress and examining them with several structural probes, it
has been possible to achieve a detailed thermodynamic and structural
characterization of the change between two distinct regions on the
liquid crystalline phase digram: a higher-density hexagonally packed
region with long-range bond orientational order in the plane
perpendicular to the average molecular direction; and a lower-density
cholesteric region with fluid-like positional order.  X-rays
scattering on highly ordered DNA arrays at high density and with the
helical axis oriented parallel to the incoming beam showed a six-fold
azimuthal modulation of the first order diffraction peak that reflects
the macroscopic bond-orientational order.  Transition to the
less-dense cholesteric phase through osmotically controlled swelling
shows the loss of this bond orientational order that had been expected
from the change in optical birefringence patterns and that is
consistent with a rapid onset of molecular positional disorder. This
change in motion was previously inferred from intermolecular force
measurements and is now confirmed by $\rm ^{31}P$ NMR.  Controlled
reversible swelling and compaction under osmotic stress, spanning a
range of densities between $\sim 120$ mg/ml to $\sim 600$ mg/ml,
allows measurement of the free energy changes throughout each phase
and at the phase transition, essential information for theories of
liquid-crystalline states.

\bigskip
\noindent
{\sl \scriptsize Submitted to PNAS (file ~/art/DNA/dnart.tex). Figures in the
directory ~/finalfig.}
\end{abstract}


\end{titlepage}


\baselineskip=15pt      
\parskip=10pt plus 1pt
\parindent=20pt


Double helical DNA has emerged as a remarkably useful material for
visualizing liquid-crystalline structures \cite{bib2} and for
measuring the packing energies associated with them. Robust DNA double
helices, of almost any monodisperse length from a few base pairs to
molecular weights $\approx 10^{9}$ can be obtained through modern
molecular biology methods and can be condensed into highly ordered
arrays easily probed by x-ray diffraction. The strong polyelectrolyte
interactions between helices can be controlled effectively by type and
concentration of the excess electrolytes.  The form of the
interhelical potential suggests that the lessons learned from
concentrated DNA arrays will have broad applications to other
seemingly unrelated physical systems such as their recently noticed
similarity to magnetic vortex arrays in type II superconductors
\cite{nelson2}.

Given the extensive investigations of the physical properties and
structure of condensed DNA phases, it is perhaps surprising that there
has not yet been a comprehensive thermodynamic characterization of DNA
mesophases under controlled solution conditions.  Following Robinson's
\cite{bib1} seminal observation of a cholesteric-like phase of long
DNA {\sl in vitro} there have appeared several studies detailing the
complexity of DNA phase behavior and its relevance for the conditions
{\sl in vivo} \cite{bib2}. The sequence of mesophases for
short-fragment DNA (146 bp $\sim$ 50 nm nucleosomal DNA) appears to
be: isotropic solution $\longrightarrow$ cholesteric $\longrightarrow$
columnar hexagonal $\longrightarrow$ hexagonal \cite{bib3}. With
biologically more relevant long-fragment DNA ($\sim 100~\rm nm$ to
$\sim \rm mm$), the sequence of phases is less well delimited and
characterized: isotropic solution $\longrightarrow$
(``precholesteric'' $\longrightarrow$) cholesteric $\longrightarrow$
columnar hexagonal $\longrightarrow$ hexagonal crystalline \cite{bib4}.
These sequences were obtained on stoichiometric mixtures of DNA, salt,
and water where there is often more than one phase present and where
neither salt nor water chemical potentials are known.

A separate line of study of the condensed phases of DNA was initiated
by Lerman \cite{bib5} through the polymer and salt induced
condensation ($\psi$ DNA) and equilibrium sedimentation
\cite{bib6} of DNA solutions.  The density of the condensed DNA was
shown to depend continuously on the concentration of condensing
polymer agent (usually polyethylene glycol, PEG) \cite{bib7}.  The use
of {\sl osmotic stress} \cite{bib8} was built on the realization that
the condensing polymer is essentially fully excluded from the DNA
phase and that, at equilibrium, the activities of the exchanging water
and salt are equal in the DNA and PEG phases \cite{bib9}.  Knowing the
osmotic pressure ($\Pi$) contribution from the excluded polymer,
measured by standard procedures as a function of its concentration,
means the osmotic pressure of the DNA is also known while all other
intensive variables such as pH and the chemical potentials of salt and
other small solutes \cite{bib8, bib9, bib10} are held fixed. Using
x-ray scattering to measure the interaxial spacing, $D_{int}$, between
double helices, this method was used successfully to elucidate a $\Pi
- D_{int}$ dependence for DNA as a function of temperature, salt type
and salt concentration \cite{bib9,bib10}.  The range of osmotic
pressures accessible through this method is substantially larger
(especially at high stress) than by the equilibrium sedimentation
approach \cite{bib6}.

Since DNA is equilibrated against a vast excess of a polymer and water
solution of known chemical potential, it is always in a single phase
at thermodynamic equilibrium.  This behavior should be contrasted with
multiple phase equilibria that usually emerge from stoichiometric
mixtures.

In this work we combine both the structural and thermodynamic
approaches to the condensed DNA phases so that structural and
dynamical parameters of DNA packing and ordering (interhelical
separation, bond orientational order parameter, $\rm ^{31}P$-NMR
spectra) are all measured concurrently with the free energy and/or its
derivatives.  We report here the structural and dynamic changes that
occur in the DNA concentration region from 120 to 600 mg/ml
corresponding to interaxial separations of 25 to 50 \AA. We show that
at lower densities (or higher spacings) DNA packing is characterized
by short range positional order, measured by x-ray diffraction, long
range cholesteric order, revealed by optical birefringence, and high
mobility  of the DNA backbone, inferred from $\rm ^{31}P$ NMR
spectrometry. At high densities (or small spacings) DNA packing is
characterized by short range positional order and long range bond
orientational order in the plane perpendicular to the average nematic
director, revealed by the azimuthal profile of the first order x-ray
diffraction peak and low mobility of the DNA backbone.

\vfill
\section{MATERIALS AND METHODS}

Wet-spun oriented samples were prepared from calf-thymus DNA
(Pharmacia) with a molecular weight of $\sim 1.6 \times 10^7$ by the
standard method \cite{bib12}.  This spinning allows controlled
production of sufficient amounts of highly oriented thin films by
spooling DNA fibres which are continuously stretched during
precipitation into an aqueous alcohol solution. Films of thickness of
$\sim 0.5~\rm mm$ and surface area between $\rm 5$ and $10~\rm mm^2$
were used in the experiments reported here.

Unoriented fibers of high MW ($\sim 1 \times 10^8$) DNA were prepared
from whole adult chicken blood (Truslow Farms, Chestertown, MD) as
described in McGhee {\sl et al.} \cite{bib11}.  This DNA was further
purified with three extractions against phenol/chloroform (50:50) and
once with chloroform alone. Then DNA was ethanol precipitated in sodium
acetate, pelleted by centrifugation, washed twice with 70\% ethanol
and dried. This DNA was used in all preparations involving unoriented
fibers.

Oriented as well as unoriented DNA fibers were equilibrated with
various solutions of PEG (20,000 MW) in 0.5 M NaCl, 10 mM Tris/ 1 mM
EDTA, pH 7 in vast excess.  Under these conditions, PEG (20,000 MW) is
completely excluded from the DNA phase for concentrations greater than
$\approx 7 \% $(w/w). The equilibration time was usually from four
days to a week. Measurements on both orientationally ordered
(wet-spun) as well as ``powder'' samples show that there is
essentially no difference in osmotic pressure vs. concentration
(interhelical spacing) dependence between the two preparations.  The
two preparations differ only in the size of the oriented domains.

X-ray diffraction was performed at $20^{\circ}$C with an Enraf-Nonius
Service Corp. (Bohemia, NY) fixed-anode FR 590 x-ray generator
equipped with image plate detectors.  Image plates were read and
digitized by a Phosphor Imager (Molecular Dynamics, CA) and processed
with NIH Image 1.55 program (W. Rasband, NIH, Bethesda, MD) modified
by us.  The position of the first order diffraction peaks
($r_{1.max}$) is obtained by radially averaging the scattering profile
around the direct beam.  Angular intensity profiles were taken at the
position of the maximum of the first order diffraction peak and were
then Fourier transformed to extract the bond orientational order
parameter ${\cal C}_6$, {\sl i.e.} the sixth order Fourier
coefficient.  If there were perfect alignment of the x-ray beam and
the average director of the oriented DNA sample , the angular
dependence of the six-fold symmetric scattering function could be
Fourier analyzed in terms of \cite{strandburg}
\begin{equation}
{\cal S}(\theta,r_{1.max}) = I_0(r_{1.max})\left[{\textstyle\frac{1}{2}} +
\sum_{n=1}^{\infty} {\cal C}_{6n}~\cos{6n(\theta - \theta_0)}\right] + I_{BG},
\label{equ0}
\end{equation}
where $I_{BG}$ is the background intensity.  Because the orientation
was only approximate there was usually a small ${\cal C}_2$ component
present in the Fourier analyzed angular profiles. We have rescaled the
value of ${\cal C}_{6}$ to correct for this .

DNA samples at various densities were sealed between microscope cover
glass and were observed under a microscope (Olympus) equipped with
crossed polarizers. The image was digitized and analyzed with NIH
Image 1.55. The ``fingerprint'' cholesteric pattern \cite{biophys}
with long fragment DNA was never as regular as is typical of short
fragment DNA. Rather long DNAs achieve oriented domains of much smaller
size.

The $\rm ^{31}P$ NMR measurements were performed on a Bruker MSL-300
spectrometer (Billerica, MA) using a high power probe with a 5 mm
solenoidal sample coil which was doubly tuned for $\rm ^{31}P$
(121.513 MHz) and protons (300.13 MHz). Gated broadband decoupled $\rm
^{31}P$ spectra were observed with a phase cycled Hahn echo sequence.
A delay time between the 90 degree pulse and 180 degree pulse of 30
microseconds was chosen. Typically 20,000 to 80,000 scans with a
recycle delay time of 1s were accumulated. Exponential linebroadening
with a linewidth of 200 Hz was used.

First moments of the NMR spectra ($M_1$) were calculated in standard
fashion according to
\begin{equation}
M_1 = \frac{\int_{-\infty}^{+\infty} f(\omega)~\omega
d\omega}{\int_{-\infty}^{+\infty} f(\omega)~d\omega},
\label{equM1}
\end{equation}
where $f(\omega)$ is the spectral intensity at the frequency $\omega$.
The frequency of the center of the spectrum, determined as half height
of the integral $\int_{-\infty}^{+\infty} f(\omega)~d\omega$, was set
to zero.

The measured dependence of the osmotic pressure of the DNA phase on
DNA concentration allows one to evaluate the reversible work done at
constant temperature, pressure and chemical potential of salt as the
system is brought from an initial (i) to a final (f)
configuration. The difference in free energy is
\begin{eqnarray}
\Delta {\cal G} &=& - \int^{V^{f}}_{V^{i}} \Pi(V_{DNA})~dV_{DNA}.
\label{equ1}
\end{eqnarray}

The excess or packing energy per unit length of the DNA helix can now
be obtained as
\begin{equation}
\frac{\Delta{\cal G}}{L} = - \sqrt{3} \int^{D_f}_{D_i}~\Pi(D)~DdD,
\label{equ4}
\end{equation}
where $D$ is the interhelical spacing assuming the DNA array is at
least locally hexagonal. Since the DNA osmotic pressure decays
exponentially at small and intermediate values of $D$, a finite
density interval is sufficient to evaluate the above integral to
satisfactory accuracy. We have taken $D_i$ corresponding to the
concentration $15 {\rm mg/ml}$ (data not shown on Fig.1), which marks
the onset of the condensed (anisotropic) DNA phase \cite{kunal}.

Since thermal fluctuations are contributing to the free energy it is
reasonable to express the calculated free energy per unit length,
$\frac{\Delta{\cal G}(D)}{L}$, in its ``natural'' units of $kT$ per
persistence length ${\cal L}_p$ ( $\approx 500$ \AA). In these units
one can write
\begin{equation}
\frac{\left(\Delta {\cal G}(D)/kT\right)}{{L}/{\cal L}_p} =
\frac{{\cal L}_p}{\zeta (D)},
\label{extra1}
\end{equation}
where ${\zeta (D)}$ is the contour length of DNA associated with kT of
packing energy in the condensed phase.

\vfill
\eject

\section{RESULTS}

\subsection{\sl Osmotic Stress Measurements}

The dependence of osmotic pressure on the concentration of the
unoriented DNA subphase has been investigated in detail \cite{bib9,
bib10, bib14}.  The corresponding interhelical spacings were obtained
by measuring the first order x-ray diffraction peak on unoriented DNA
samples assuming local hexagonal packing symmetry.  This assumption
was verified experimentally in the high density region (I) (see Fig.1)
through the existence of weak higher order reflections and now by
observing well developed six-fold symmetric bond orientational order
(see section 2.2).

Similar measurements were performed on oriented samples that show the
same interaxial spacing (or density) dependence on $\Pi$ as the
unoriented samples (see Fig. 1) and thus have the same free energy,
within experimental error.  There are two distinct regions in the
${\Pi} - D$ curve. In the high pressure regime, the interhelical
distance does not depend on the salt concentration .  The forces
between helices in this region were interpreted as resulting from
water - mediated structural forces \cite{bib9}.  At lower pressures a
sensitivity of D to salt concentration is clearly discernible. The
effective decay length for the interhelical interactions, however, is
about twice the predicted Debye screening length \cite{bib15} for salt
concentrations $<$ 1.0 M, where electrostatic interactions are not
overwhelmed by hydration forces.  The two scaling regimes of the
osmotic pressure are separated by a narrow crossover region in the
${\Pi} - D$ curve at about $ 32 - 34$~\AA.

\subsection{\sl Packing Symmetry}

The two regimes in the osmotic pressure curve are also clearly evident
in the qualitative characteristics of the X-ray diffraction on
oriented samples (see Fig. 2).  For oriented samples of DNA in the
high osmotic pressure regime (I) the cross section of the first order
interaxial diffraction peak with the DNA helical axis oriented
parallel to the incoming beam is a circular ring with six-fold
modulation in the intensity which clearly reflects the six-fold
symmetric long range bond orientational order of the underlying DNA
lattice Fig.2 (inset). Azimuthal modulation of the first order
diffraction peak at close DNA spacings has been observed previously in
neutron diffraction studies \cite{bib13} with fibers of NaDNA and
LiDNA at low excess salt content. As the osmotic pressure is lowered
the six-fold modulation of the first order diffraction peak disappears
and is unobservable below the transition, 32-34 \AA, region (see inset
Fig.2) in the $\Pi - D$ curve. For spacings less than 35 ~\AA~ the
changes in the six-fold modulation of the diffraction peak were
reversible. However, once the bond orientational order is lost, it
cannot be regained by simply increasing the osmotic pressure. The
subsequent chain entanglement due to the looser nature of the packing
in this low pressure phase apparently precludes the reestablishment of
long range bond orientational order. The details of the first order
diffraction peak are irretrievably lost leading to a circular powder
pattern.

The details of the azimuthal profile of the diffraction pattern were
independent of the X-ray beam size up to cross sectional areas on the
order of $\sim \rm mm^2$. The bond orientational order thus appears to
be of very long range indeed. The translational order, on the other
hand, estimated crudely from the radial linewidth of the first order
diffraction peak \cite{chaikin} and extremely weak higher order
reflections (J. R\" adler, personal communication), appears to be of a
much shorter range, on the order of several lattice spacings.

To quantify this change in orientational bond order, we have measured
the azimuthal intensity distribution of the first order diffraction
peak and extracted the corresponding Fourier coefficients shown in
Fig.2.  Generally the Fourier spectra showed pronounced peaks for
${\cal C}_n$ with $n = 0 ~{\rm and}~ 6$, with typically a small, but
discernible additional contribution from ${\cal C}_2$, most probably
reflecting a slight misorientation of the x-ray beam direction and the
average director of the oriented DNA sample. The extracted ${\cal
C}_6$ coefficients, that are also corrected for misalignment, shows a
gradual loss of lateral bond orientational order as the DNA density
passes from the high to low osmotic pressure regimes .

The nature of the low osmotic pressure phase can be further
ascertained by polarized light microscopy which clearly reveals the
existence of a ``fingerprint'' texture characteristic of a cholesteric
phase \cite{Livolant}. Though the pitch of the cholesteric phase
varies with density of the DNA phase in the vicinity of I
$\longrightarrow$ II transition, we were unable to quantify this
accurately because the orientational domain sizes were, in general,
small. Due to the high molecular weight of the DNA, the samples could
not be manipulated by an applied external orienting magnetic field to
increase the domain size.

\subsection{\sl Phosphate Backbone Dynamics}

An earlier analysis of the $\Pi - D$ curve suggested that there was a
relatively sudden change in lattice fluctuations, inferred from
changes in x-ray scattering peak widths, within the 32-34 \AA\
transition region, \cite{bib10}. This change in motion can now be seen
very clearly in the $\rm ^{31}P$ NMR spectra. The insert in Fig. 3
shows two $\rm ^{31}P$ NMR spectra - one within the high pressure
regime and one in the low pressure, cholesteric phase - that clearly
demonstrate a symmetry change in the effective tensor of chemical
shift.  While any quantitative relation between the values of the
effective tensor of chemical shift, the spectral first moment, and the
details of the molecular motions is highly model dependent, it is
clear that there is a qualitative difference in the DNA dynamics
between the two pressure regimes.  We have quantified this change by
analyzing the first moments of the $\rm ^{31}P$ NMR spectra, shown in
Fig.3. If there are no other processes contributing to resonance
broadening and since osmotically equilibrated samples are
monophasic, the observed increase of the first moment as the
interaxial spacing decreases is due to a decreased mobility of DNA
helices.

The molecules are obviously immobilized to a substantial degree in the
high pressure phase though the spectral first moment does decrease
somewhat , see Fig.3, as the 32-34 \AA\ transition region is
approached.  In the low pressure region, the phosphate mobility
appears to be significantly greater. The difference in the spectral
first moments between the two pressure regimes suggests a drastic
increase of motional amplitudes for the cholesteric phase but the
mobility appears not to change substantially with density within this
phase.

Typical principal values of the chemical shift tensor extracted from
spectra at high applied osmotic stress are $\sigma_{xx} \approx -60$
ppm, $\sigma_{yy} \approx -5$ ppm, and $\sigma_{zz} \approx 65$ ppm.
Comparable values for essentially completely immobilized dry DNA, are
$-83$ ppm, $-22$ ppm and $110$ ppm (measured relative to 85\%~
phosphoric acid as a standard) \cite{shindo}. The effective tensor for
DNA in the high pressure regime shows that phosphate motions are quite
restricted.  No fast rotation around one axis is present because this
would have resulted in a tensor with axial symmetry.

\vfill
\eject

\section{DISCUSSION}

\subsection{\sl Structure and Dynamics}

This study, together with earlier measurements of intermolecular
forces \cite{bib9,bib10}, presents a departure from the usual
gravimetric method of sample preparation.  By bringing ordered phases
into equilibrium with large "reservoirs" of salt-plus-polymer
solutions rather than by making stoichiometric mixtures of salt, water
and DNA, it is possible to set all the intensive thermodynamic
variables associated with the resulting single liquid-crystalline phase.

These simultaneous measurements of the structure, motion and
thermodynamic functions of DNA phases have focused on high density DNA
phases (with interhelical spacings between about 25 and 55 \AA) at one
ionic strength (0.5 M NaCl).  This density region extends from $\sim
120~\rm mg/ml$ to $\sim 600~\frac{mg}{ml}$. At lower densities there
is a transition to a cholesteric phase from one of the (presumably)
blue phases \cite{amelie}, while at higher densities there is a
transition into a three dimensional crystal with a simultaneous B
$\longrightarrow$ A transition in DNA conformation. For the long
fragment DNA investigated here, the isotropic $\longrightarrow$
anisotropic transition is still quite remote ($\sim 10~\rm mg/ml$
\cite{kunal}).

The structural, dynamic, and osmotic stress data presented here are
all consistent with the existence of two different DNA phases
separated by a transition region at a DNA density of $\sim 320 - 360
{}~\rm mg/ml$, corresponding to interhelical spacings of $\sim 32 -
34$~\AA. Previous work on short fragment ($146~{\rm bp} \sim 500$~\AA~
long) DNA \cite{bib3} also gave clear evidence for the existence of a
series of structurally distinct regions as a function of DNA
concentration.  The transition from a cholesteric to a 2D-hexagonal
phase for short fragment DNA was observed at $\sim 32$~\AA. Remarkably
the $\sim 32$~\AA~ interaxial spacing is also close to the spacing
from which ${\rm Mn}^{2+}$ or ${\rm Co}^{3+}$-DNA collapses in a first
order transition under osmotic stress \cite{don}. Is this a distance
at which the details of the chiral double-helical structure come to be
sensed in molecular interaction?

What these experiments do not show clearly is the nature and the order
of the transition between bond orientationally ordered and cholesteric
phases. There is no detectable discontinuity ( the accuracy of the
measurement of the interhelical spacing in this regime of DNA
densities is $\sim 1$ \AA ) in the $\Pi$ {\sl vs.} interaxial
separation curve that is seen when DNA makes a clear first-order
transition \cite{don}.  This should not be taken as definitive
evidence, however, that the transition is second order.  An extremely
narrow phase coexistence window could simply be a property of polymers
in liquid crystalline mesophases \cite{edwards}. The accuracy of the
azimuthal scans of the first order diffraction peak as well as the
first moment of the $\rm ^{31}P$ NMR spectra also precludes a
definitive measure of the order of the transition.

\subsection{\sl Free Energy and Intermolecular Forces}

The "osmotic stress" exerted by the excluded polymer is the rate of
change of free energy with change in the amount of solution in the DNA
phase, i.e., $\Pi = -{\partial G}/{\partial V_{DNA}}$. By integrating
the osmotic pressure curve one thus obtains the change in the system
free energy, Eq.\ref{equ4}. In the insert to Fig. 1 we have plotted
this free energy as a function of molecular separation.  It
is given in thermal units of kT per persistence length (see
Eq.\ref{extra1}), and spans a wide range of energy scales, from about
kT per 2.5~\AA~ at $\log(\Pi) \sim 8~{\rm dynes/cm^2}$ to about kT
per 100~\AA~ at $\log(\Pi) \sim 6~{\rm dynes/cm^2}$.

Previous work \cite{bib14} has established that forces in the high
pressure regime are dominated by exponentially decaying hydration
interactions with a decay length $\lambda \sim 3 - 4$~\AA~ that is
basically independent of the ionic strength. In the low pressure
regime, the interaxial spacing dependence on osmotic stress is also
exponential, but the effective decay length is about twice the
expected Debye decay length (at least for salt concentrations between
{}~ 0.2 and $\sim 0.8$~ M). The enhanced decay length and a rescaling of
the strength of the interactions between DNA helices in this regime of
DNA densities was shown to be due to the progressive onset of
conformational disorder characterized by the fluctuations in the mean
position of the molecules along the average director \cite{bib14, flu}
and deduced from the width of the interhelical x-ray scattering
peaks. The switch between fluctuation enhanced forces and bare
potentials was not found to be gradual, but rather quite abrupt as the
DNA density passed the $\sim 340~\rm mg/ml$ limit \cite{bib10},
correlating nicely with emergence of longitudinal order between
helices seen in the studies of Livolant {\sl et al.} \cite{bib3} on
short fragment DNA, as well as with the onset of lateral bond
orientational order and broadening of the phosphate NMR peak reported
here.

The fluctuation-enhanced effective interactions observed in DNA arrays
have the same origin as the effective interactions in smectic
arrays. They are due to the interplay between conformational
fluctuations and bare short range potentials \cite{bib15}.  The
clearly emerging enhancement of electrostatic decay length to about
twice the Debye length, not yet so easily seen in lipid bilayer
smectic arrays, could be connected with the different dimensionalities
of the two systems (2D periodicity {\sl vs.} 1D periodicity).

\subsection{\sl Perspectives and Directions}

Molecular interactions in DNA arrays, extracted from the measured
osmotic pressure of the array, are expected \cite{bib10} to vary with
the interaxial spacing $D$ as $\sim K_0(D/\lambda)$ , where $ K_0(x)$
is the modified Bessel function with asymptotic behavior $ K_0(x)
\approx (\pi /2x)^{1/2}~e^{-x}$, with a decay length $\lambda$
dependent on the salt concentration \cite{bib14}.  In this respect, as
noted by Nelson \cite{nelson2}, the interactions between helices in
condensed DNA mesophases are formally and surprisingly closely related
to the interactions between magnetic vortex lines in flux-line
lattices of high-$\rm T_c$ superconductors which, apart from the lack
of hard core repulsions, share the same form of interaction potential.

The existence of a line (polymer) hexatic phase, intermediate between a line
crystal and a line liquid, was hypothesized by Marchetti and Nelson
\cite{nelson1} specifically for the case of magnetic flux-line
lattices. It appears that the bond ordered DNA phase (region I)
described above is perhaps this type of intermediate phase. The
transition from a line hexatic phase in DNA into one of the possible
less ordered phases is complicated by the presence of chiral coupling
in the molecular interactions at lower densities, leading to the
cholesteric, not a line liquid, phase. The occurrence of line hexatic
between the cholesteric and the crystalline (A-form DNA) phases makes
it difficult to compare directly with existing theoretical
predictions. Its existence nevertheless introduces a new possible
scenario into the melting sequence of ordered polyelectrolyte arrays.

To say that DNA provides an opportunity to learn about liquid-crystals
is not to say that it has already given clear answers to basic
questions. What is the nature of the transition from a phase with well
developed six-fold symmetric bond orientational order to a a skewed,
cholesteric phase when the molecules are allowed to move apart?  Why
does this change in symmetry couple with the molecular motions that
cause extra interaxial separation \cite{bib10}? What is the nature of
molecular packing in the long polymer cholesteric phase compared to
the more common twisted nematic phases of shorter molecules? These
combined structural studies \cite{bib3}, osmotic stress measurements
of free energies, and x-ray \& NMR probes of molecular disorder and
motion now provide a direction and an opportunity for further
development of systematic theoretical analyses.

\vfill
\eject

\vfill
\eject

\noindent
{\Large \bf Figure captions}

{\bf Figure 1.}~{\bf Osmotic pressure of DNA, stressed by solutions of
PEG (20,000 MW) at different concentrations, as a function of the
interhelical spacing for 0.5 M NaCl DNA}, for wet-spun, highly
oriented DNA $\bullet$ and for condensed unoriented DNA $\Diamond$. No
significant difference between the macroscopically oriented and
unoriented samples is observed over the osmotic stress region
investigated. The upper continuous curve represents a fit to the
experimental data with bare hydration forces at high pressures (region
I) and fluctuation enhanced screened Coulombic interactions in the low
pressures region (II), while the lower curve represents a hypothetical
osmotic pressure dependence if the underlying forces were a simple sum
of bare hydration and screened Coulomb interactions, neglecting the
contribution from DNA conformational disorder. The dotted vertical
region represents the phase boundary between the bond orientationally
ordered and the cholesteric phases. A DNA density scale in mg/ml is
also given in order to facilitate comparison with previous work. The
insert presents the free energy in units of kT per persistence length
($\sim 500$ \AA~ or $\sim 150$ base pairs). $\frac{{\cal
L}_p}{\zeta(D)} = 1$ corresponds to one persistence length per kT
while $100$ corresponds to one hundredth of a persistence length (or
$\sim 1.5$ base pairs) per kT of interaction energy.

\bigskip

{\bf Figure 2.}~{\bf The dependence of the bond orientational order
parameter ${\cal C}_6$ on the interaxial spacing in the region of the
high-pressure to low-pressure transition}. The vanishing of the long
range bond order is clearly visible.  Six-fold modulation in the first
order diffraction peak seen directly in the scattering patterns
(inset, for left to right, $\log(\Pi) = 7.80, 7.51, 6.795~{\rm
dynes/cm^2}$), gradually weakens and disappears below the transition
region (shown by the dotted area) identified in the $\Pi - D$ shown in
Fig. 1. This loss of bond orientational order is reversible provided
the interhelical spacing does not exceed the transition region ($\sim
35$~\AA).

\bigskip

{\bf Figure 3.}~{\bf The first moment $M_1$ of the $\rm ^{31}P$ NMR
spectra as a function of the interhelical separation $D$} measured on
the same (unoriented) samples as with x-ray scattering.  There is a
qualitative change in the shape of the NMR spectra (see inset) as the
system goes through the high-pressure to low-pressure transition
region (the dotted area).  The inset shows two NMR spectra: at $\Pi =
7.98~{\rm dynes/cm^2}$ corresponding to interhelical spacing of
$26$~\AA~ (broad spectrum) and at $\Pi = 6.119~{\rm dynes/cm^2}$
corresponding to interhelical spacing of $46$~\AA~ (narrow spectrum)
and ionic strength 0.5 M NaCl.  Much more motional freedom of DNA
phosphates is evident at low pressures than high.  The change in the
behavior of the spectral linewidth at the transition region is quite
abrupt.

\noindent

\end{document}